\documentclass[12pt]{article}
\usepackage{amsmath}
\usepackage{amssymb}
\usepackage{amsthm}
\usepackage{graphicx}
\usepackage{psfrag}
\usepackage[hang, nooneline]{subfigure}
\usepackage{color}

\usepackage{soul} 
\usepackage{cite}

\usepackage{amssymb} 
\usepackage{amsfonts} 
\usepackage{amsmath} 
\usepackage{slashed}

\usepackage{egothic}%
\usepackage{pgothic}%
\usepackage{yfonts}%

 \usepackage{yhmath} 

\newcommand{\we}{\wedge}

\newcommand{\der}{\partial}

\newcommand{\inn}{\hspace*{2pt}\raisebox{-1pt}{\rule{6pt}{.3pt}\hspace*
{0pt}\rule{.3pt}{8pt}\hspace*{3pt}}}

\newcommand{\beq}{\begin{equation}}
\newcommand{\eeq}{\end{equation}}
\newcommand{\beqa}{\begin{eqnarray}}
\newcommand{\eeqa}{\end{eqnarray}}
\newcommand{\nn}{\nonumber}

\newcommand{\pbr}[2]{ \{ \hspace*{-2.6pt} [ #1 , #2\hspace*{1.4 pt} ] 
\hspace*{-2.6pt} \} }
\newcommand{\bx}{{\mathbf{x}}}

\newcommand{\BH}{{\bf H}} 
\newcommand{\BS}{{\bf S}}

\author{ 
\vspace*{1ex} \\
Monika E. Pietrzyk$^{1}$,  C\'ecile Barbachoux$^{2}$ and Joseph Kouneiher$^{2}$
\\ \small $^{1}$ Mathematics and Physical Sciences, University of Exeter, 
 EX4 4QL Exeter, UK
\\ 
\small {\!\!\!}${\!\!\!}^{2}$  
Sciences and Technologies Department, 
\\ \small C\^ote d'Azur University/INSPE,
06000 Nice, France 
}

\date{} 

\title{
\Large \bf 
The relation between the canonical Hamilton-Jacobi equation and the 
covariant Hamilton-Jacobi equation for Maxwell's electrodynamics 
}

\begin{document}
\maketitle
\begin{abstract}
The aim of this paper is to understand the relation between the canonical Hamilton-Jacobi equation for Maxwell's electrodynamics, which is an equation in variational derivatives for a functional of field configurations, and the covariant (De Donder-Weyl) Hamilton-Jacobi equation, which is a partial derivative equation on a finite dimensional space of vector potentials and spacetime coordinates. We show that the procedure of spacetime splitting applied to the latter allows us to reproduce both the canonical Hamilton-Jacobi equation and the Gauss law constraint in the Hamilton-Jacobi form without a recourse to the canonical Hamiltonian analysis. Our consideration may help to analise the quasiclassical limit of the connection between the standard quantization in field theory  based on the canonical Hamiltonian formalism with a preferred time dimension and the precanonical quantization that uses the De Donder-Weyl Hamiltonian formulation where space and time dimensions treated equally. 
\end{abstract}

\newpage

\section{Introduction}

A covariant generalization of the Hamiltonian formalism to field theory is possible in two distinct ways. The first way is the canonical formalism 
which can be formulated covariantly on the infinite-dimensional space of solutions (see, e.g., \cite{garcia,witten,crnkovic,marsden,kijowski,binz})  and requires a global foliation of spacetime into space-like leaves. Its underlying structure is the symplectic two-form on the infinite dimensional space of solutions or initial data. The second way (see, e.g., \cite{dedonder,rund,kastrup}) treats all spacetime variables equally 
as ``independent variables" of the variational problem that defines a field theory. The underlying structures are known as multisymplectic \cite{kij} 
or polysymplectic \cite{gunther,ikbr1,ikbr2,sardan},  and they are represented by forms of degree $(D+1)$ 
in the spacetime dimension $D$ or related geometrical constructions on a finite-dimensional polymomentum analogue of the phase space 
(see \cite{deleon-book} for a review and comparison). Both approaches lead to their own generalizations of the Hamilton's canonical equations from mechanics to field theory, their own regularity conditions of the respective Legendre transformations, and the corresponding analysis of constraints (compare, e.g., \cite{regge} with \cite{deleon-constr,kan-dirac,rr}). They also have their own respective analogs of Poisson brackets which are defined, in the canonical formalism, for functionals of field configurations and, in the polysymplectic realization of the spacetime symmetric formalism \cite{ikbr1,ikbr2}, for horizontal differential forms on a finite dimensional bundle \cite{ikbr3,ik5,khbr1,khbr2,khbr3}, and they lead to different approaches to quantization of fields known as canonical quantization (see, e.g., \cite{schweber}) and precanonical quantization put forward in 
\cite{ik5,ik1,ik2,ik3,ik4,ik5e}. 
The approach of precanonical quantization has been applied recently to the problems of quantum gravity and quantum gauge theory 
\cite{ikm1,ikm2,ikm3,ikm4,ikv1,ikv2,ikv3,ikv4,ikv5,iky1,iky2,iky3,iktp1,iktp2}, 
and to the problem of cosmological constant \cite{ikcosm1,ikcosm2}. 
The relation between precanonical quantization and the standard quantum field theory based on canonical quantization has been discussed in \cite{iks1,iks2,iksc1,iksc2,iksc3}.

Both covariant generalizations of the Hamiltonian formalism to field theory lead to their own analogues of the Hamilton-Jacobi formulation (compare, e.g., \cite{weiss-hj,deleon-hj} with \cite{dedonder,weyl,rund,vonrieth,kastrup}). Our recent papers,  where a covariant (De Donder-Weyl) Hamilton-Jacobi equation is derived for the teleparallel equivalent of general relativity \cite{mcec} and for Maxwell's theory in Palatini-like formulation \cite{ours23}, we continued this line of research. The relation between  the canonical Hamilton-Jacobi formulation (on the infinite dimensional space of configuration of fields), 
 which is an equation in variational derivatives, and the covariant ((De Donder-Weyl) Hamilton-Jacobi formulation (on a finite dimensional space whose coordinates are field variables and spacetime variables), which is a partial differential equation, is not yet understood in general. 
However, in a few particular examples, it was demonstrated how the canonical Hamilton-Jacobi equation in variational derivatives 
is derived from the covariant (De Donder-Weyl) Hamilton-Jacobi partial differential equation by applying a procedure of decomposition of spacetime into the space and time and by constructing the action (or eikonal) functional of the initial data from the De Donder-Weyl eikonal functions restricted to an  initial field configuration \cite{ik-pla,nikolic,riahi}.     

The aim of this paper is to extend this relation to the simplest gauge field, i.e. the electromagnetic field. To this purpose, the canonical Hamilton-Jacobi formulation for Maxwell field theory is outlined in Section 1 and then, in Section 2, we show how it is derived from the covariant (De Donder-Weyl) Hamilton-Jacobi equation known from our previous work \cite{ours23} and earlier papers \cite{dedonder,vonrieth,horava} by a spacetime split and a restriction to the subspace of initial data and integration over it. Our terminilogy and notation follow our previous paper \cite{ours23}.

\section{Canonical Hamilton-Jacobi equation} 

In the canonical Hamiltonian formalism we split the spacetime variables $x^\mu$ into the space variables $\bx$ and the time variable $x^0 = t $. Then, canonical momenta derived 
from the Lagrangian 
 \beq \label{lagr40}
 L=-\frac14 F^{\mu\nu}F_{\mu\nu} , 
 \eeq 
 where 
 \beq
 F_{\mu\nu} := \der_\mu A_\nu - \der_\nu A_\mu , 
 \eeq
are 
\beq
p_{A_0}(\bx) = 0,  \quad  p_{A_i}(\bx) = - F^{0i}(\bx) , 
\eeq
and the canonical Hamiltonian has the form 
\beq
\BH = \int\!d\bx\ \left( \frac12 (F^{0i})^2 + \frac14 F^{ij}F_{ij}
+ F^{0i}\der_i A_0 \right) . 
\eeq
The integration by parts in the last term leads to 
\beq
 - \int\!d\bx\ A_0(\bx) \der_i F^{i0} (\bx)
\eeq 
so that the non-dynamical $A_0$ appears as the Lagrange multiplier and  fixes the Gauss law constraint 
\beq
\der_i F^{i0} (\bx) = 0 
\eeq
The canonical Hamilton-Jacobi equation is formulated for a functional of field configurations at a fixed moment of time $\BS([A^i(\bx)], t)$ such that 
\beq 
p_{A_i}(\bx) = \frac{\delta \BS}{\delta {A_i}(\bx)}
\eeq  
and it thus takes the form 
\beq \label{hj44}
\der_t \BS = \int\!d\bx\ \left( \frac12  \left( \frac{\delta \BS}{\delta {A_i}(\bx)}\right)^2 
+ \frac14 F^{ij}F_{ij} \right) . 
\eeq
The Gauss law in Hamilton-Jacobi formulation has the form 
\beq \label{hj45}
\der_i \frac{\delta \BS}{\delta A_i(\bx)} = 0 
\eeq 
The solutions of the Maxwell equations are related to the solutions of (\ref{hj44}), 
(\ref{hj45}) by means of the embedding condition 
\beq 
\frac{\delta \BS}{\delta {A_i}(\bx)} = F^{i0}(\bx)
\eeq
An earlier appearence of this form of the Hamilton-Jacobi formulation is  found in Max Born's paper \cite{born}
and also in the context of the 
David Bohm's causal interpretation in quantum field theory \cite{kaloy1,kaloy2}. 

\section{Canonical Hamilton-Jacobi from the 
covariant \\Hamilton-Jacobi equation} 

 The covariant De Donder--Weyl (DDW) HJ equation for Maxwell field theory 
 has the form \cite{ours23,dedonder,vonrieth,horava,zajac}
 \beq \label{e35}
\der_\mu S^\mu - \frac{1}{4} \frac{\der S^\mu}{\der A_\nu }\frac{\der S_\mu}{\der A^\nu } .
=0 , 
\eeq
where $S^\mu$ are functions of the field components $A_\nu$ and spacetime coordinates $x^\mu$: $ S^\mu = S^\mu (A,x)$.  
The solutions of Maxwell's equatons $A_\mu(x)$ are related to the solutions of the DDW HJ quation (\ref{e35})  
by using the embedding condition 
\cite{ours23} 
\beq  \label{embed}
F^{\mu\nu} (x) = - \left.\left( \frac{\der S^{[\mu}}{\der A_{\nu ]} }\right) \right\vert_\sigma  , 
\eeq
where $|_\sigma$ means a restriction to a particular field configuration $\sigma$  which is a section $A_\mu (x)$ in the total space 
of the bundle of the electromagnetic potentials $A_\mu$ over the spacetime with the coordinates $x^\mu$. 
If the spacetime split is performed such that $x^\mu := (t, \bx)$ then the configuration $\sigma$ is determined by the initial data $\Sigma$
 at an initial moment of time: $A_\mu(\bx)$.

We would like  to understand the relation between the canonical HJ formulation in terms of the functional $\BS([A(\bx)], t)$ and the covariant 
 DDW HJ  formulation that uses several functions $S^\mu (A,x)$. 
 Let us start from the assumption that (cf. \cite{ik-pla,riahi}) 
\beq  \label{dwhjmax}
\BS = \int_\Sigma S^\mu (A_\mu, x^\mu)|_{{}_\Sigma}\ \upsilon_\mu =  \int\! d\bx\  S^0 (A_\mu(\bx),\bx,t)  ,
\eeq
where $\upsilon_\mu := {i}_{\der_\mu} (dx^0\we dx^1\we... \we dx^{D-1})$ and $\Sigma$ represents the surface of initial data 
$ A_\nu = A_\nu (\bx)$ at a fixed moment of time. 
In the following we denote $S^\mu|_{{}_\Sigma} = S^\mu (A_\nu(\bx),\bx,t)$ simply as $S^\mu$. 

The time derivative of\, $\BS$ can be obtained from the covariant De Donder-Weyl (DDW) HJ equation (\ref{e35}). 
Namely,  
\begin{align}\label{e39}
\der_t \BS =\int\!  d\bx\ \der_t S^0  = \int\!  d\bx\ \left( -\der_i S^i
+ \frac{1}{4} \frac{\der S^0}{\der A_\nu }\frac{\der S_0}{\der A^\nu }
+ \frac{1}{4} \frac{\der S^i}{\der A_\nu }\frac{\der S_i }{\der A^\nu } \right)  . 
\end{align}
The first term can be 
rewritten as 
\beq \label{e41}
 \int\!  d\bx\ \der_i S^i =  \int d\bx\ \left( \frac{ d S^i}{dx^i} 
- \der_i A_\mu (\bx) \frac{\der S^i}{\der A_\mu } \right) 
= - \int\! d\bx\  \der_i A_\mu (\bx) \frac{\der S^i}{\der A_\mu }  , 
\eeq
where the total divergence whose integral is vanishing is given by 
\beq
\frac{ d S^i}{dx^i} = \der_i S^i + \der_i A_\mu(\bx) \frac{\der S^i}{\der A_\mu} .
\eeq

Using the constraints, we can write the right hand side of (\ref{e41}) as 
\beq
  \int\!d\bx\ \left ( -  \der_i A_0 (\bx) \frac{\der S^0}{\der A_i  }  
 + \der_i A_j (\bx)  \frac{\der S^{[i}}{\der A_{j]} } \right )  .
\eeq
Integrating by parts the first term and using the embedding condition 
(\ref{embed}) in the second term 
we obtain 
\beq
\int\! d\bx\ \left ( A_0 (\bx) \frac{d}{dx^i} \frac{\der S^0}{\der A_i }  
-  \frac12 F_{ij}F^{ij}  \right ) .
\eeq

In the spacetime with the signature $(+,-,-,-)$ which we use here,  the second term in (\ref{e39}) 
transforms into 
\beq
- \frac{1}{4} \int\! d\bx\  \frac{\der S^0}{\der A_i }\frac{\der S_0}{\der A_i }  .
\eeq 
Note that 
\beq 
\frac{\der S^0}{\der A_0} = 0  
\eeq 
 due to the constraint 
\beq \label{e36}
p^{(\mu }_{A_{\nu)}} = \frac{\der S^{(\mu}}{\der A_{\nu)} } 
=0 , 
\eeq 
which follows from the definition of polymomenta from the Lagrangian (\ref{lagr40}) \cite{ours23}. 

Using the constraints (\ref{e36}) and the relation between polymomenta and field strengths, which follows 
from the Lagrangian (\ref{lagr40}), namely,  
\beq
p^{\mu }_{A_{\nu}} = - F^{\mu\nu}, 
\eeq
and the embedding condition of the DDW HJ formulation (\ref{embed}), 
 the last term in (\ref{e39})  can be transformed as follows 
\begin{align}
& \frac{1}{4}\int d\bx\ \left( - \left( \frac{\der S^i}{\der A_0 }\right)^2 
+ \frac{\der S^i}{\der A_j }\frac{\der S_i }{\der A^j}\right ) \\
& =   \int d\bx\ \left( - \frac{1}{4}\left( \frac{\der S^0}{\der A_i }\right)^2 
 + \frac{1}{4} F^{ij} F_{ij} \right ) . 
\end{align}

By taking note of the fact that 
\beq
\frac{\delta \BS}{\delta A_i(\bx)} = \frac{\der S^0 }{\der A_i }(A(\bx),\bx)
\eeq 
we obtain from  (\ref{e39})
\begin{align}
\label{e57} 
\!\der_t \BS &= \int\! d\bx \left( A_0 (\bx) \frac{d}{dx^i} \frac{\der S^0}{\der A_i  (\bx)}  
- \frac12 (F_{ij})^2 
- \frac14 \left( \frac{\delta \BS}{\der A_i(\bx)}\right)^2 
- \frac{1}{4} \left( \frac{\delta \BS}{\der A_i(\bx)}\right)^2
 +\frac14  (F_{ij})^2  
\right)  \nn \\
&  \\
&= - \int d\bx \left(\frac12 \left(\frac{\delta \BS}{\delta A_{i}(\bx)} \right)^2 
+ \frac14 F_{ij}(\bx)F^{ij}(\bx)  
- A_0 (\bx) \frac{d}{dx^i} \frac{\delta \BS}{\delta A_i  (\bx)}  \right)  .
\nn
\end{align}

 Thus, without any recourse to the procedures of canonical Hamiltonian formalism, 
 we obtain the canonical Hamilton-Jacobi equation 
\beq 
\der_t \BS + \int d\bx \left( \frac12 \left(\frac{\delta \BS}{\delta A_{i}(\bx)} \right)^2 
+\frac14 F_{ij} F^{ij} 
\right) =0
\eeq  
and the Gauss law 
in the Hamilton-Jacobi form 
\beq
 \frac{d}{dx^i}\frac{\delta \BS}{\delta A_i  (\bx)}  = 0 
\eeq 
by applying the spacetime split and integration over the initial data to the covariant De Donder-Weyl HJ equation (\ref{e35}). 
The Gauss law constraint  arises with the Lagrange multiplier $A_0(\bx)$, and the latter decouples from the dynamics. 
(cf. \cite{regge,iky3}).

\section{Conclusion}

Our discussion shows that the formula (\ref{dwhjmax}) conjectured in \cite{ik-pla},  
which establishes the connection between the De Donder-Weyl HJ functions $S^\mu$ and the canonical 
HJ action functional $\BS$, is also valid for Maxwell field theory. 
It allows us to derive the canonical variational derivative HJ equation with a distinguished time variable 
from the partial derivative De Donder-Weyl HJ equation that treats all spacetime variables equally. 
Quite unexpectedly, the procedure of spacetime splitting leads to the automatic emergence of the Gauss law constraint from the 
covariant De Donder-Weyl formulation. It is similar to the authomatic emergence of the quantum Gauss law 
from the derivation of the canonical functional derivative Sch\"odinger equation from the precanonical 
 Sch\"odinger equation in quantum gauge theory in \cite{iky3}. One can conjecture that by choosing different 
$(D-1)-$dimensional foliations and then integrating over their leaves, one can derive canonical HJ equations in different gauges. 
The clarification of the relation between the canonical HJ theory of fields and the  De Donder-Weyl HJ theory may also 
help to understand the quasiclassical limit of the Schr\"odinger equation for quantum fields derived by precanonical quantization 
(see the References cited in the Introduction) and a possible Bohmian interpretation of it similar to that studied recently 
in \cite{derak}.  It is also interesting to extend our analysis to more general non-abelian gauge fields (cf. \cite{iky3,molgado1}) and to try to 
obtain the canonical Hamilton-Jacobi formulation of the teleparallel equivalent of general relativity (TEGR) by applying the procedure 
of this paper to our recent work \cite{mcec} on the covariant HJ formulation of TEGR instead of using the cumbersome canonical Hamiltonian 
treatment of the system with a hierarchy of constraints.

\end{document}